\documentclass[preprint2]{aastex7}

\begin{document}

\title{Hidden Transits: TOI-2285 b is a Warmer sub-Neptune Likely with a Super-Earth Companion}

\author[0000-0002-4909-5763]{Akihiko Fukui}
\affiliation{Komaba Institute for Science, The University of Tokyo, 3-8-1 Komaba, Meguro, Tokyo 153-8902, Japan}
%\affiliation{Instituto de Astrof\'isica de Canarias, V\'ia L\'actea s/n, E-38205 La Laguna, Tenerife, Spain}
\email{afukui@g.ecc.u-tokyo.ac.jp}



\begin{abstract}

TOI-2285\,b is a sub-Neptune-sized planet orbiting a nearby M dwarf, discovered through the TESS photometric survey and ground-based follow-up observations. The planet was initially reported to have an orbital period of 27.27\,d, making it one of the lowest temperature sub-Neptunes transiting a bright M dwarf. However, additional TESS data reveal that its true orbital period is 13.64\,d, half the original value, resulting in a warmer equilibrium temperature (358\,K) than previously estimated (284\,K). The misidentification likely resulted from the low signal-to-noise ratio of individual transit signals and the limited number of transits observed by TESS at that time. This case highlights the importance of carefully considering harmonic solutions for similar cases. The additional TESS data also reveal another planetary candidate with an orbital period of 9.67\,d and a radius of 1.5\,$R_\oplus$, requiring validation in future studies.

\end{abstract}

%\keywords{Exoplanet systems (484), Super Earths (1655), Transit photometry (1709)}

\section{Introduction} \label{sec:intro}

Although more than two thousands of sub-Neptune-sized ($1.6 < R_p/R_\oplus < 4$) planets have been discovered through transit surveys, only a limited number ($\sim$200) receive low irradiation (instellation flux $\lesssim$ 5 $S_\oplus$). These planets are of particular interest because their lower atmospheric mass-loss rate allows them to retain more of their primordial atmospheres.
Those orbiting bright host stars, in particular, must offer valuable insights into the formation and evolution of super-Earths and sub-Neptunes through detailed follow-up observations.

TOI-2285\,b is a sub-Neptune-sized (1.8\,$R_\oplus$) planet orbiting a nearby (42.4\,pc) and bright ($J = 9.86$\,mag) M dwarf \citep[][hereafter AF22]{2022PASJ...74L...1F}, which was initially identified as a planetary candidate from three sectors of TESS survey data \citep{2015JATIS...1a4003R} and subsequently validated by ground-based follow-up observations. The planet was originally reported to have an orbital period of 27.27\,d, which corresponds to an instellation flux of 1.54\,$S_\oplus$ and equilibrium temperature of 284\,K. This made TOI-2285\,b one of the few known low-irradiated sub-Neptunes transiting nearby M dwarfs.
Since then, TESS has reobserved TOI-2285 in six additional sectors. I revisit the system by analyzing these new data to refine the physical parameters of TOI-2285\,b and search for signals of additional planets.

\section{Analysis and Results} \label{sec:analysis}

TESS has observed TOI-2285 in nine sectors (Sectors 16, 17, 24, 56, 57, 76, 77, 83, and 84) all with a two-minute cadence. The Pre-search Data Conditioning Simple Aperture Photometry (PDCSAP) light curves, produced by the TESS Science Processing Operations Center \citep[SPOC,][]{2016SPIE.9913E..3EJ}, were retrieved from the Mikulski Archive for Space Telescopes (MAST)\,\footnote{\dataset[https://doi.org/10.17909/he2t-4s41]{https://doi.org/10.17909/he2t-4s41}} (Figure \ref{fig1} (a)). The light curves were detrended using the Savitzky-Golay filter, and transit signals were searched using the Transit Least Squares (TLS) algorithm \citep{2019A&A...623A..39H} for periods ranging from 0.1 to 100 days. The strongest signal appeared at 13.64\,d with a signal detection efficiency (SDE) of 54.2, which exceeds the nominal detection threshold of 9 and is significantly higher than for the period previously reported of 27.27\,d (top panel of Figure \ref{fig1} (b)). The phase-folded light curves at 13.64\,d showed consistent transit depths and durations between odd- and even-number transits. No significant transit signal was found at a phase of 0.5, ruling out the possibility of a quarter of the original period. I therefore conclude that the true orbital period of TOI-2285\,b is 13.64\,d instead of 27.27\,d. 
 
After the transit signals of TOI-2285\,b were masked, the TLS analysis was repeated, revealing a signal of another planetary candidate at a period of 9.67\,d with an SDE of 18 (bottom panel of Figure \ref{fig1} (b)). The same signal has also been identified by the dedicated pipeline of SPOC \citep[][and references therein]{2019PASP..131b4506L} from Sectors 16--84, which is to be released as TOI-2285.02 (J. M. Jenkins, private communication). No further signals were found from the subsequent analysis.

I analyzed the undetrended PDCSAP light curves with transit and systematic trend models, following the procedure of \cite{2021AJ....162..167F}.
The fitting parameters include impact parameter $b$, log of the scaled semi-major axis $\log (a/R_s)$, planet-to-star radius ratio $R_p/R_s$, two coefficients of the quadratic limb-darkening law $u_1$ and $u_2$, and mid times of individual transits $T_c$. Uniform priors were applied to all parameters except $T_c$, $u_1$, and $u_2$, for which Gaussian priors were applied.

The derived median values and 1$\sigma$ uncertainties are as follows: $R_p/R_s = 0.0348\ ^{+0.0021}_{-0.0016}$, $b = 0.44\ ^{+0.32}_{-0.31}$, and $a/R_s = 43.1\ ^{+5.9}_{-11.8}$ for TOI-2285\,b, and $R_p/R_s = 0.0271\ ^{+0.0030}_{-0.0022}$, $b = 0.55\ ^{+0.32}_{-0.38}$, and $a/R_s = 42\ ^{+12}_{-17}$ for TOI-2285.02.
The best-fit transit models are shown in Figure \ref{fig1} (c).
A linear regression to the measured $T_c$ for each object gives the transit ephemerides in ${\rm BJD_{TDB}}$ of $T_{\rm c}(E) = 2458747.1831 (29) + 13.635109 (31) \times E$ and $T_{\rm c}(E) = 2458747.6202 (26) + 9.673341 (19) \times E$ for TOI-2285\,b and TOI-2285.02, respectively, where $E$ is the transit epoch and the numbers in parentheses represent the last two digits of 1$\sigma$ uncertainties. 
Adopting the values of the stellar parameter values from AF22, I derive planetary radii of $1.77\ ^{+0.12}_{-0.09}\ R_\oplus$ and $1.37\ ^{+0.16}_{-0.12}\ R_\oplus$, instellations of $3.91\ ^{+0.36}_{-0.33}\ S_\oplus$ and $6.17\ ^{+0.57}_{-0.52}\ S_\oplus$, and equilibrium temperatures (assuming a bond albedo of 0.3) of $358 \pm 8$\,K and $402 \pm 9$\,K for TOI-2285\,b and TOI-2285.02, respectively.

\section{Discussions} \label{sec:conclusion}

The refined instellation and equilibrium temperature indicate that the planet is more irradiated and warmer than previously estimated ($1.54 \pm 0.14 \ S_\oplus$ and $284 \pm 6$\,K, respectively), yet it remains one of the limited number ($\sim$20) of low-irradiated ($<$5\,$S_\oplus$) sub-Neptunes transiting bright ($J < 11$\,mag) M dwarfs.

\begin{figure*}
    \centering
    \gridline{\fig{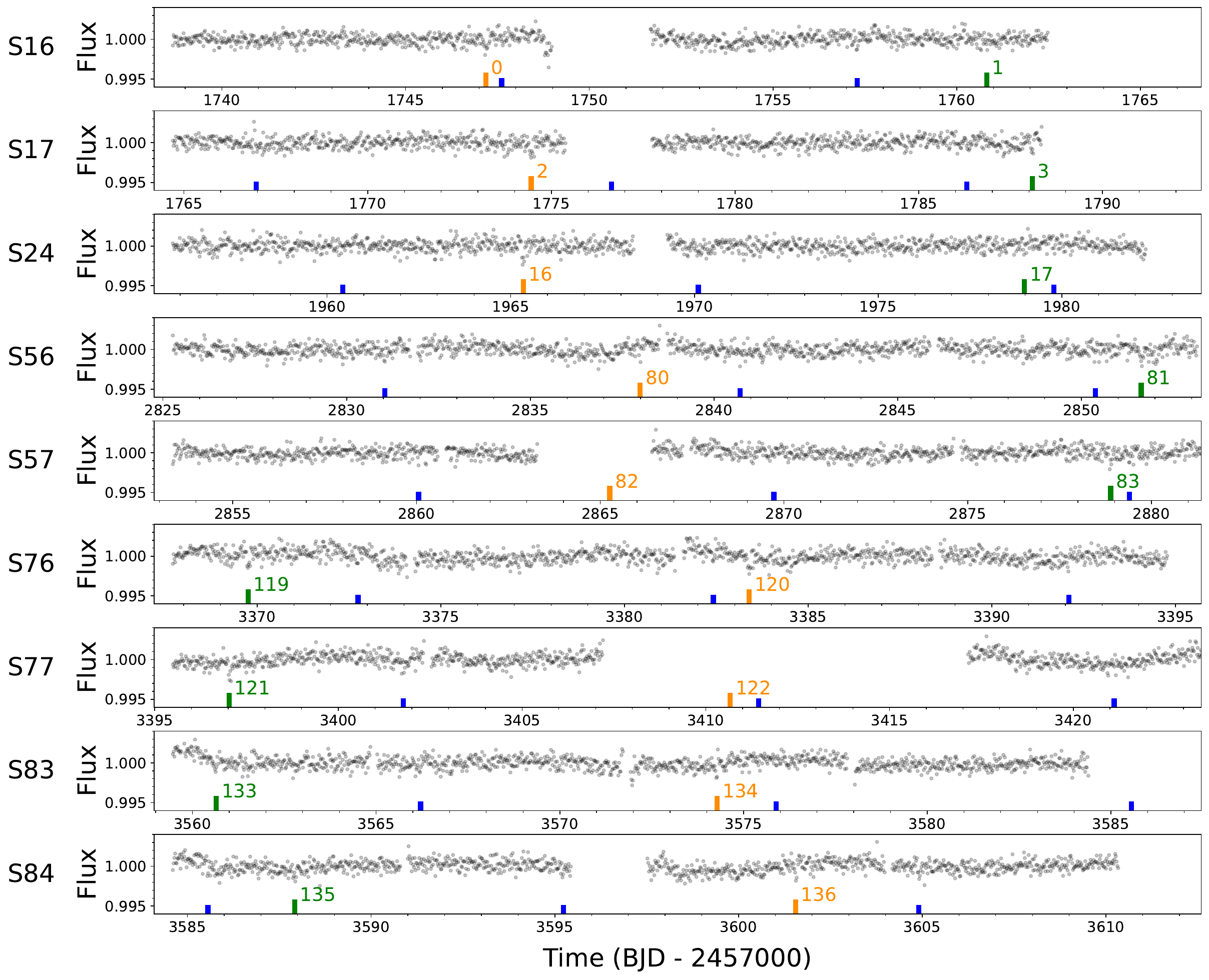}{0.87\textwidth}{(a)}}
    \gridline{\fig{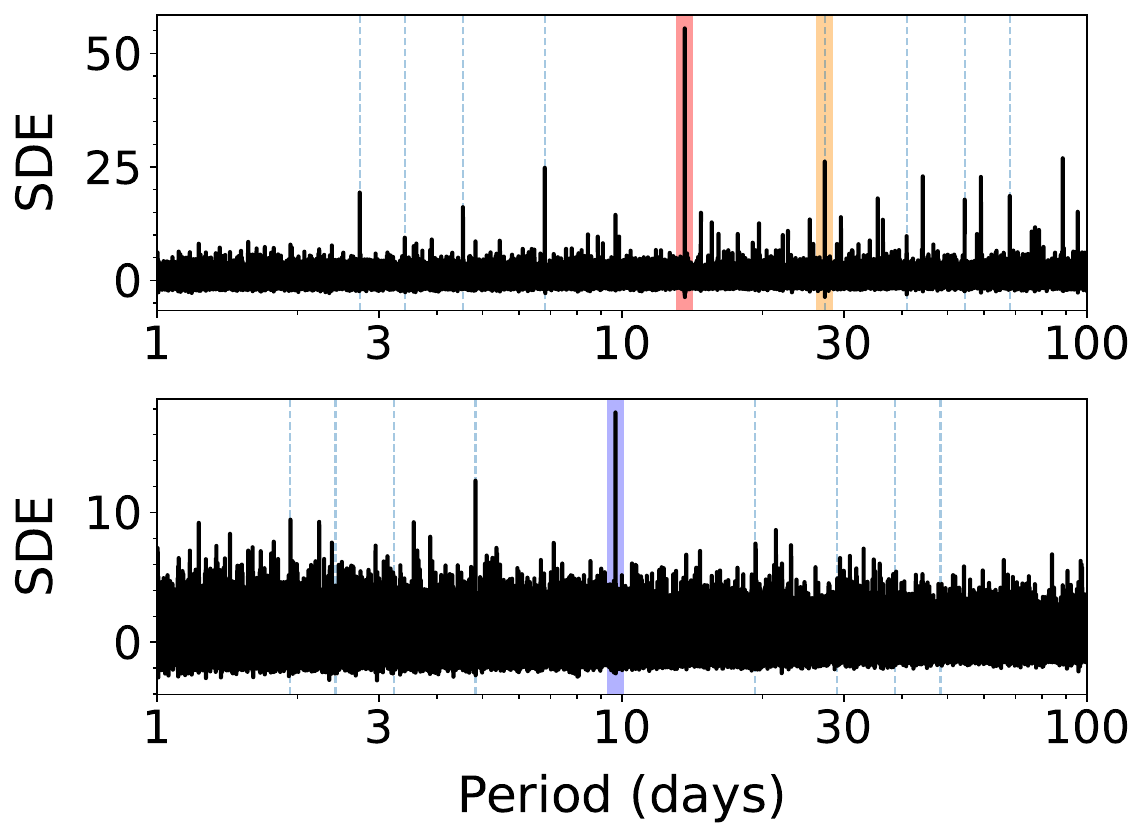}{0.4\textwidth}{(b)}
              \fig{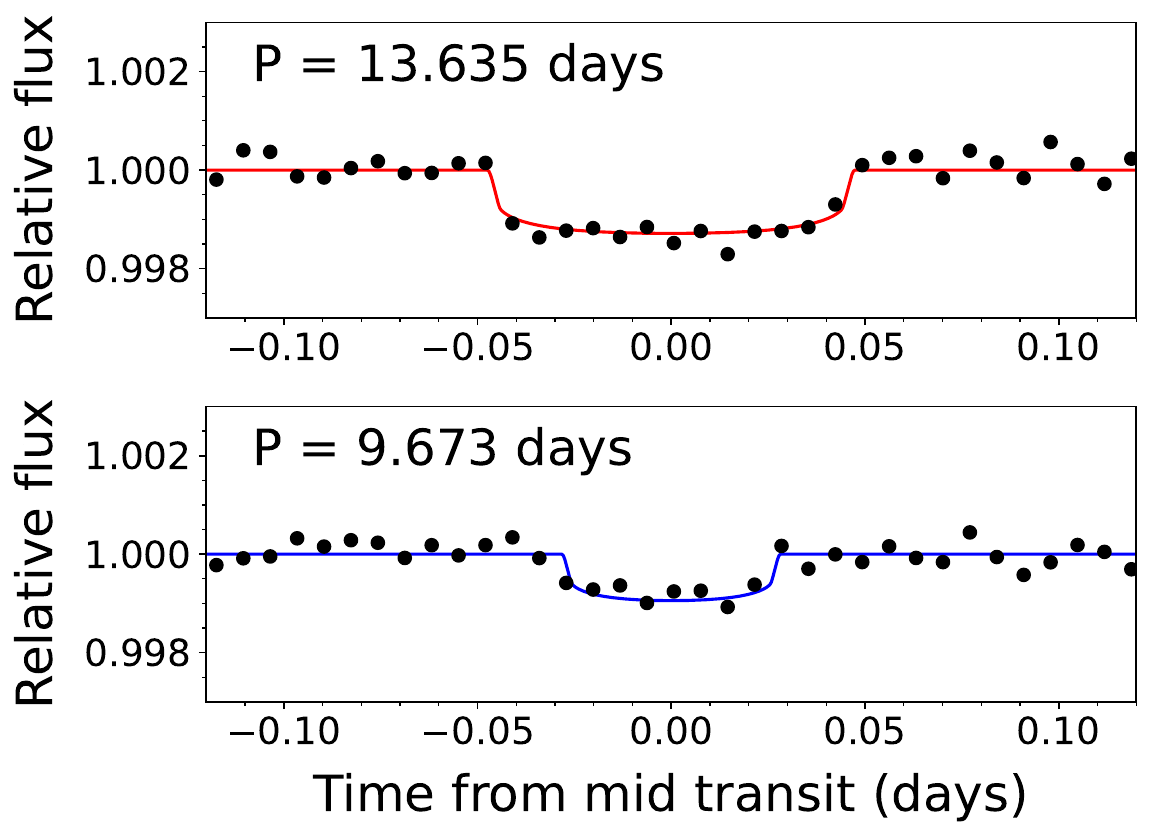}{0.4\textwidth}{(c)}}
    \caption{(a) The sector-to-sector undetrended PDCSAP light curves of TOI-2285 binned by 30 minutes. The sector number is indicated on the left. The locations of even- and odd-number transits of TOI-2285\,b are marked by orange and green, respectively, with transit epochs labeled. The locations of the transits of TOI-2285.02 are marked by blue. (b) The power spectra obtained from the first (top) and second (bottom) TLS analyses. The initial and refined orbital periods of TOI-2285\,b are marked by orange and red bold lines, respectively, while the signal of TOI-2285.02 is marked by blue bold line. The blue dashed lines indicate the harmonics of the highest peaks. (c) Detrended, phase-folded, and 10-minutes-binned PDCSAP light curves for TOI-2285\,b (top) and TOI-2285.02 (bottom). The solid lines are the best-fit transit models. }
    \label{fig1}
\end{figure*}

The orbital period reported by AF22 was identified from the first three sectors by the SPOC's pipeline, with all verification tests passed.
This initial misidentification likely resulted from the low signal-to-noise ratio (SNR) of the individual transit signals and the small number of transits. Although six transits actually exist in the first three sectors, the pipeline detected only the signals associated with even-number transits (transit epochs 0, 2, and 16; see Figure \ref{fig1} (a)). The odd-number transits (transit epochs 1, 3, and 17) are less evident than the even-number transits, probably due to statistical fluctuations and/or systematic noises, leading the pipeline to overlook the odd-number transits. Note that the SPOC's pipeline has recently recovered the true orbital period from Sectors 16--77, as reported in the Threshold Crossing Event (TCE) summary report released on February 19, 2025 \footnote{Available in MAST (observation ID: tess2019199201929-s0014-s0078-0000000329148988).}.
The same issue could happen in other TESS planetary candidates when the orbital period is relatively long, the data length is limited, and/or the SNR of individual transits is small. 
The case of TOI-2285\,b highlights the importance of carefully considering harmonic solutions under such conditions.

The additional TESS data have also revealed a candidate signal of a super-Earth-sized planet, warranting validation in future studies.

\begin{acknowledgements}

I thank J. M. Jenkins for his useful information and comments. 
I acknowledge the use of public TESS data from pipelines at the TESS Science Office and at the TESS SPOC. 
This work is partly supported by JSPS KAKENHI Grant Number JP24K00689.

\end{acknowledgements}

%\bibliography{ref}{}
\bibliographystyle{aasjournal}

\end{document}